\documentclass[journal]{IEEEtran}
\IEEEoverridecommandlockouts

\usepackage[ruled,linesnumbered]{algorithm2e}
\usepackage{cite}
\usepackage{amsmath,amssymb,amsfonts}
\usepackage{algorithmic}
\usepackage{graphicx}
\usepackage{bm}
\usepackage{float}
\usepackage{subfigure}
\usepackage{textcomp}
\usepackage{xcolor}
\usepackage{amsthm}
\usepackage{breqn}
\usepackage{pifont}

\usepackage[backref]{hyperref}

\def\BibTeX{{\rm B\kern-.05em{\sc i\kern-.025em b}\kern-.08em
    T\kern-.1667em\lower.7ex\hbox{E}\kern-.125emX}}

\begin{document}

\title{Privacy Attacks and Defenses for Digital Twin Migrations in Vehicular Metaverses}

\author{Xiaofeng Luo, Jinbo Wen, Jiawen Kang*, \textit{Senior Member, IEEE}, Jiangtian Nie, Zehui Xiong, Yang Zhang, Zhaohui Yang, Shengli Xie, \textit{Fellow, IEEE}
\thanks{
X. Luo, J. Kang, and S. Xie are with the School of Automation, Guangdong University of Technology, China (e-mail: gdutxiaofengluo@163.com; kavinkang@gdut.edu.cn; shlxie@gdut.edu.cn). J. Wen and Y. Zhang are with the College of Computer Science and Technology, Nanjing University of Aeronautics and Astronautics, China (e-mail: jinbo1608@163.com; yangzhang@nuaa.edu.cn). J. Nie is with the School of Computer Science and Engineering, Nanyang Technological University, Singapore (e-mail: jnie001@e.ntu.edu.sg). Z. Xiong is with the Pillar of Information Systems Technology and Design, Singapore University of Technology and Design, Singapore (e-mail: zehui\_xiong@sutd.edu.sg). Z. Yang is with the College of Information Science and Electronic Engineering, Zhejiang University, China (e-mail: yang\_zhaohui@zju.edu.cn).

(\textit{*Corresponding author: Jiawen Kang})
}
}

\maketitle

\begin{abstract}
The gradual fusion of intelligent transportation systems with metaverse technologies is giving rise to vehicular metaverses, which blend virtual spaces with physical space. As indispensable components for vehicular metaverses, Vehicular Twins (VTs) are digital replicas of Vehicular Metaverse Users (VMUs) and facilitate customized metaverse services to VMUs. VTs are established and maintained in RoadSide Units (RSUs) with sufficient computing and storage resources. Due to the limited communication coverage of RSUs and the high mobility of VMUs, VTs need to be migrated among RSUs to ensure real-time and seamless services for VMUs. However, during VT migrations, physical-virtual synchronization and massive communications among VTs may cause identity and location privacy disclosures of VMUs and VTs. In this article, we study privacy issues and the corresponding defenses for VT migrations in vehicular metaverses. We first present four kinds of specific privacy attacks during VT migrations. Then, we propose a VMU-VT dual pseudonym scheme and a synchronous pseudonym change framework to defend against these attacks. Additionally, we evaluate average privacy entropy for pseudonym changes and optimize the number of pseudonym distribution based on inventory theory. Numerical results show that the average utility of VMUs under our proposed schemes is 33.8$\%$ higher than that under the equal distribution scheme, demonstrating the superiority of our schemes.
\end{abstract}

\begin{IEEEkeywords}
Metaverse, vehicular twin, privacy protection, migration, inventory theory.
\end{IEEEkeywords}

\IEEEpeerreviewmaketitle

\begin{figure*}[t]
\centering{\includegraphics[width=0.95\textwidth]{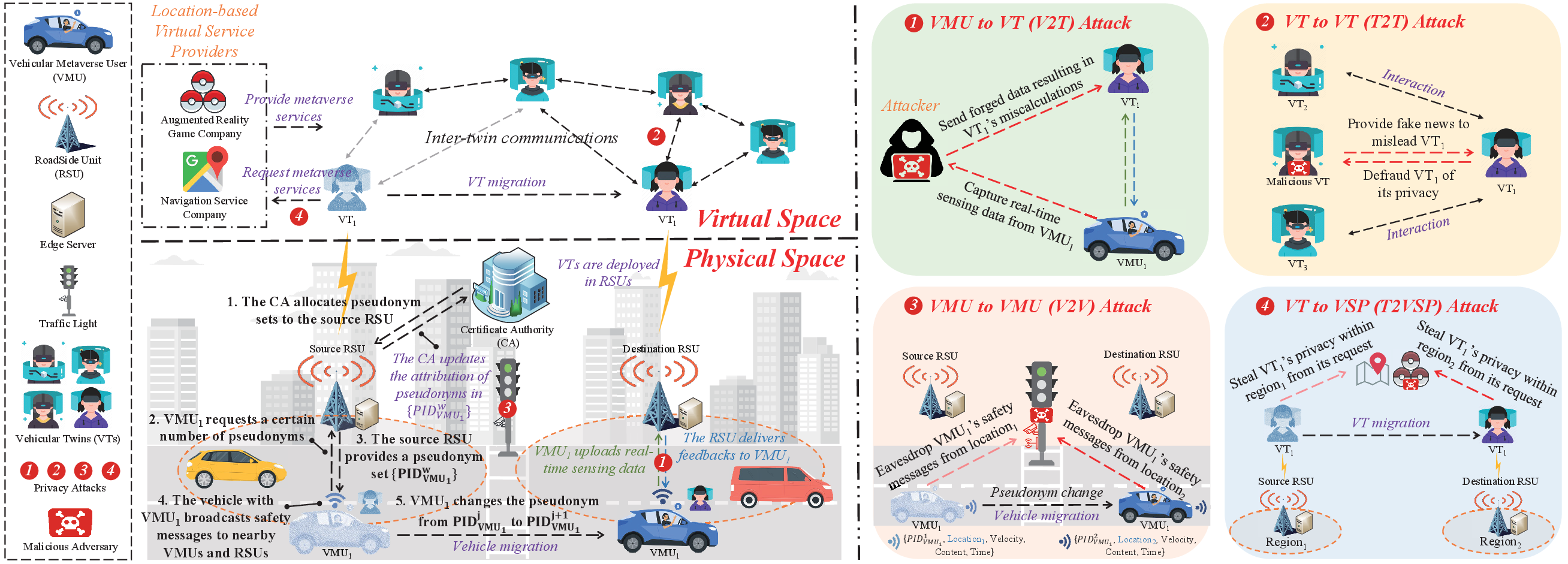}}
\caption{An overview of privacy attacks during VT migrations in vehicular metaverses. The left part introduces the VT migration process in vehicular metaverses. The right part provides a detailed description of four kinds of privacy attacks during VT migrations.}\label{system}
\end{figure*}

\section{Introduction}
The emergence of advanced technologies, such as Web 3.0 and metaverses, has spurred an increased interest in intelligent transportation systems from both industry and academia\cite{zhou2022vetaverse}. Especially, the synergy between transportation systems and metaverses has given rise to the concept of vehicular metaverses. The vehicular metaverse is regarded as a blended immersive realm that integrates extended reality technologies and real-time vehicular data to provide diverse and personalized in-vehicle services for Vehicular Metaverse Users (VMUs) (i.e., drivers and passengers within vehicles)\cite{zhou2022vetaverse}. Based on the role of digital twins in metaverses, which are virtual representations of real-world entities, Vehicular Twins (VTs), highly accurate and large-scale digital replicas that cover the lifecycle of the vehicle and VMUs, serve as the foundation for vehicular metaverses, enabling emerging vehicular applications such as Augmented Reality (AR) navigation\cite{Jinbo}. To achieve physical-virtual synchronization, VTs are continuously updated with real-time sensing data from surrounding environments, making vehicular metaverses autonomous and sustainable\cite{zhou2022vetaverse}.

Since the construction and maintenance of VTs require significant computing resources at the network edge\cite{yu2022bi}, VMUs normally offload large-scale rendering tasks of creating and updating VTs to nearby edge servers (e.g., RoadSide Units (RSUs))\cite{Jinbo}. However, due to the limited communication coverage of RSUs and the high mobility of VMUs, VTs need to be migrated among RSUs along the moving trajectories of their associated VMUs to provide real-time and uninterrupted metaverse services. Traditionally, in vehicular metaverses, vehicles communicate with others and periodically broadcast safety messages (including pseudonyms and location information) to ensure driving security, with the pseudonyms serving as temporary identifiers for identity anonymization\cite{kang2016location}. Despite the use of pseudonyms, privacy and security threats remain a big concern for VT migrations in vehicular metaverses. To be specific, since VTs constantly interact with other VTs and request immersive services from Virtual Service Providers (VSPs) in virtual spaces\cite{xu2022fulldive}, attackers can observe the location information of VTs before and after migrations among RSUs. Combined with safety messages in the physical space, they can establish mapping relationships between VMUs and VTs. In this case, the identity and location privacy of VMUs and VTs may be leaked and easily exploited by attackers for malicious purposes, potentially compromising the security of VMUs. Therefore, it is necessary to study privacy issues and develop efficient defense schemes for VT migrations in vehicular metaverses.

Some efforts have been conducted to investigate privacy issues in the metaverse\cite{wang2023DTsurvey,xu2022fulldive}. For example, the authors in \cite{wang2023DTsurvey} comprehensively summarized the privacy and security threats in the Internet of digital twins from several perspectives, including data-related, communication-related, and privacy threats, and then discussed key research challenges to defend them. In addition, they examined effective countermeasures against these threats and assessed their feasibility in the Internet of digital twins. However, the existing work ignores potential security and privacy threats during digital twin migrations in metaverses, especially in vehicular metaverses.

To address the aforementioned challenges, we aim to investigate the privacy issues and develop reliable defense strategies for VT migrations in vehicular metaverses. \emph{To the best of our knowledge, this is the first research work to study the privacy issues and defenses for VT migrations in vehicular metaverses}. Our contributions are summarized as follows:

\begin{itemize}
    \item We introduce the VT migration process in vehicular metaverses and present four kinds of new attacks that can compromise the identity and location privacy of VMUs and VTs during VT migrations.
    \item To defend against these attacks, we propose an efficient VMU-VT dual pseudonym scheme, in which we use VT pseudonyms to achieve identity anonymization of VT communications in virtual spaces.
    \item Furthermore, to combat a special threat resulting from asynchronous pseudonym changes between VMUs and VTs, we further propose a synchronous pseudonym change framework to resolve the privacy leakage issues during VT migrations.
    \item We derive average privacy entropy to quantify the increased degree of privacy protection after pseudonym changes, and then utilize inventory theory to optimize the number of pseudonym distribution. Numerical results demonstrate that our proposed schemes can effectively ensure the privacy preservation of VMUs during VT migrations in vehicular metaverses.
\end{itemize}

\section{Privacy Attacks for Vehicular Twin Migrations and Corresponding Defenses}
In this section, we first introduce the VT migration process in vehicular metaverses and study potential privacy attacks. Then, we present our defense schemes to counter these attacks.

\subsection{Vehicular Twin Migrations}
We first introduce four key components of the vehicular metaverse as follows:

\begin{itemize}
    \item \textbf{Vehicular Twins (VTs):} As highly accurate and large-scale digital replicas of vehicles and VMUs, VTs can analyze the status of vehicles and VMUs and facilitate vehicle decision making through real-time interactions between virtual spaces and the physical space \cite{Jinbo}. Moreover, VTs can interact with other VTs for data sharing, helping VMUs obtain global environment information\cite{wang2023DTsurvey}. Therefore, VTs make the vehicular metaverses autonomous and durable.
    \item \textbf{Vehicular Metaverse Users (VMUs):} By using lightweight devices like Head-Mounted Displays (HMDs), VMUs can access vehicular metaverses to obtain immersive and lower-latency metaverse services, such as AR navigation and virtual games\cite{zhou2022vetaverse,Jinbo}. For real-time updates of VTs in the virtual space, VMUs collect real-time sensing data (e.g., real-time vehicular status and traffic condition information) from surrounding environments by vehicular sensors\cite{Jinbo}.
    \item \textbf{Roadside Units (RSUs):} RSUs are generally treated as vehicular communication devices mounted along the roadside. Empowered by edge computing technology\cite{yu2022bi}, RSUs have sufficient computing and storage resources to construct VTs and deliver ultra-reliable and low-latency metaverse services to VMUs \cite{Jinbo}. To ensure seamless and immersive experiences for VMUs, VTs in the virtual space are migrated from the source RSUs to the destination RSUs along driving trajectories of corresponding VMUs in the physical space. In addition, RSUs can serve as pseudonym caching stations responsible for the storage, management, and distribution of pseudonyms\cite{johar2021proofBC}.
    \item \textbf{Virtual Service Providers (VSPs):} VSPs are third-party entities (e.g., companies) that can provide high-quality metaverse services for VTs \cite{xu2022fulldive}. For instance, VSPs can provide location-based metaverse services for VTs based on their personalized demands, such as AR games and navigation. In this case, VSPs would collect the private information of VTs, including their previous contents of interest and current locations of corresponding VMUs.
\end{itemize}

As shown in Fig. \ref{system}, vehicles periodically broadcast safety messages to ensure driving security during VT migrations. The safety message generally includes the VMU pseudonym and real-time sensing data from surrounding environments. When communicating with other VMUs, VMUs leverage pseudonyms to conceal their true identities and constantly change their pseudonyms through driving to ensure privacy protection in vehicular metaverses.

For the convenience of explanation, we take $VMU_1$ as an illustration. We consider that the Certificate Authority (CA) and RSUs are trusted entities in line with the assumption in\cite{lyu2023enabling,yu2015mixgroup}. The CA maintained by government agencies first generates a specific number of pseudonyms and allocates them to RSUs in the form of pseudonym sets\cite{kang2016location,xu2021efficient}. When VMU pseudonyms are running out, $VMU_1$ requests a specific number of pseudonyms from the nearest RSU. Then, the RSU distributes a pseudonym set \{$PID_{VMU_1}^w$\}, where $w$ represents the number of pseudonyms and $PID_{VMU_1}^j$ is one of the pseudonyms in \{$PID_{VMU_1}^w$\}. To ensure driving security, the vehicle of $VMU_1$ broadcasts safety messages \{\textit{Pseudonym}, \textit{Location}, \textit{Velocity}, \textit{Content}, \textit{Time}\} to nearby vehicles and RSUs \cite{kang2016location}. After driving for a while, $VMU_1$ decides to change the current pseudonym $PID_{VMU_1}^j$ to avoid being tracked by attackers. The $VMU_1$ selects a new pseudonym $PID_{VMU_1}^{j+1}$ from the pseudonym set \{$PID_{VMU_1}^w$\} and changes its pseudonym from $PID_{VMU_1}^j$ to $PID_{VMU_1}^{j+1}$. Finally, $VMU_1$ uses the changed pseudonym to communicate with other VMUs and repeats this process throughout the journey, thus reducing the risk of identity leakage.

\subsection{Privacy Attacks for Vehicular Twin Migrations}
Despite the application of VMU pseudonyms, VT migrations can still arouse several unprecedented privacy concerns in vehicular metaverses. As shown in the right part of Fig. \ref{system}, we specifically present four kinds of privacy attacks during VT migrations as follows:

\begin{figure*}[t]
\centering{\includegraphics[width=0.95\textwidth]{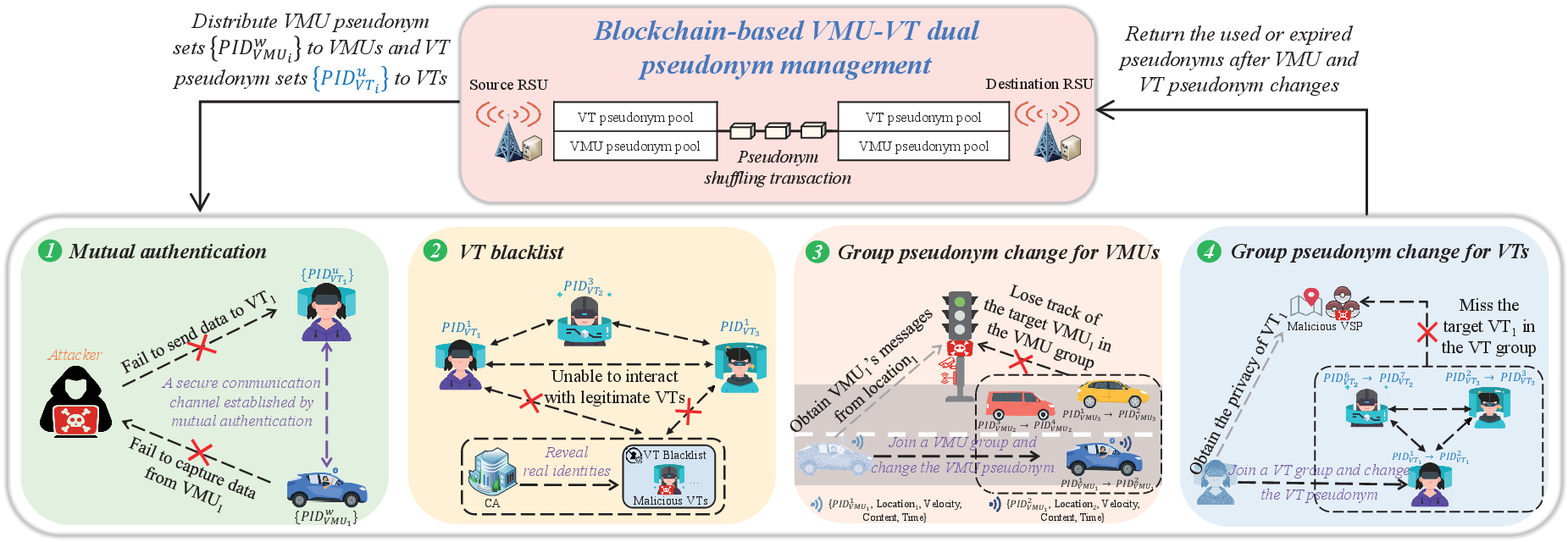}}
\caption{A VMU-VT dual pseudonym scheme consisting of four modules for defending against four kinds of privacy attacks illustrated in Fig. \ref{system}. Note that blockchain technology is utilized to securely manage both VMU and VT pseudonyms by recording pseudonym shuffling transactions.}
\label{scheme}
\end{figure*}

\begin{itemize}
    \item \textbf{\emph{VMU to VT (V2T) Attack}:} The \textit{V2T attack} occurs during authentication between VMUs and VTs. Attackers exploit system flaws to gain unauthorized access to transmitted data between VMUs and VTs \cite{siwakoti2023IoTattack}, posing a serious threat to the identity privacy of VMUs. Specifically, the vehicle of $VMU_1$ collects real-time sensing data, e.g., the distance from the vehicle in front, and uploads collected data to $VT_1$ to forecast $VMU_1$'s actions through a trained machine learning model\cite{yu2022bi}. However, the attackers lurking in the open transmission channel can capture and tamper with these sensing data, e.g., modifying the safe following distance from $100\:\rm{m}$ to $10\:\rm{m}$, and then send the forged data to the virtual space resulting in the miscalculation of $VT_1$. In this case, $VT_1$ will send incorrect feedback to misguide the driving decision making of $VMU_1$, which probably leads to a severe traffic accident.
    
    \item \textbf{\emph{VT to VT (T2T) Attack}:} The \textit{T2T attack} is launched by the purposeful interaction of malicious VTs during inter-twin communications (i.e., communications between VTs) \cite{wang2023DTsurvey} in the virtual space, which can lead to unwitting identity privacy leakages of legitimate VTs. Given the low-expense nature of socializing in vehicular metaverses, malicious VTs are prone to interact with legitimate VTs and steal their private information\cite{wang2023DTsurvey} to conduct purposeful activities such as precisely advertising or even committing crimes with stolen identities. To be specific, a malicious VT may impersonate an intimate friend of target VMUs to deliberately interact with target VTs, thereby defrauding their sensitive privacy illegally. Besides, malicious VTs can provide fake news to satisfy their own needs. For instance, a malicious VT broadcasts a non-existent accident to $VT_1$, compelling $VMU_1$ to take an alternate route. 
    
    \item \textbf{\textit{VMU to VMU (V2V) Attack}:} Ever-changing VMU pseudonyms are used to protect privacy when vehicles broadcast safety messages. However, attackers can still leverage radio equipment installed on roadside infrastructures (e.g., traffic lights) to launch \textit{V2V attacks} \cite{yu2015mixgroup}. Specifically, an attacker first eavesdrops safety messages of $VMU_1$ from $Location_1$ and then eavesdrops safety messages again from $Location_2$ after a period of $VMU_1$ migration. Although the pseudonym of $VMU_1$ has been changed from $PID_{VMU_1}^{1}$ to $PID_{VMU_1}^{2}$, the attacker can still track $VMU_1$ by analyzing similar features of safety messages (e.g, velocity and time) from two geographically adjacent locations. Therefore, the V2V attack causes a sharp decline in the location privacy of VMUs.
    
    \item \textbf{\textit{VT to VSP (T2VSP) Attack}:} To immerse themselves in location-based metaverse services (e.g., AR video games), VTs would supply location-related information to VSPs (e.g., AR game companies). This process may trigger \textit{T2VSP attacks} launched by malicious VSPs. The communication coverage of the source RSU and the destination RSU are denoted as $Region_1$ and $Region_2$, respectively. The malicious VSPs first steal the privacy of $VT_1$ deployed in the source RSU within $Region_1$. When $VMU_1$ leaves from $Region_1$ to $Region_2$, $VT_1$ is migrated from the source RSU to the destination RSU correspondingly. Afterwards, the malicious VSP can obtain the privacy of target $VT_1$ within $Region_2$. By analyzing the spatio-temporal factors of information (e.g., driving directions and timestamp) from different regions, the malicious VSP can locate and track the target $VT_1$, indicating that the T2VSP attack seriously violates the location privacy of VTs.
\end{itemize}

If multiple colluded attackers not only eavesdrop safety messages including pseudonyms and locations of target VMUs but also obtain identity information of VTs, the attackers can infer mapping relationships between VMUs and their associated VTs \cite{kang2016location}. Under this circumstance, the attackers can keep track of the target VMUs due to the immutability of VT identities, causing critical damage to the identity and location privacy of both VMUs and VTs. Furthermore, these external attackers can exploit the privacy to conduct targeted advertising, or even use stolen identities to commit crimes to avoid liability. Consequently, it is necessary to design a reliable defense scheme against the four kinds of attacks enumerated above to safeguard the privacy of both VMUs and VTs.

\subsection{Proposed Defenses: A VMU-VT Dual Pseudonym Scheme} \label{solution}

To defend against the aforementioned attacks, we design a VMU-VT dual pseudonym scheme, in which the VT pseudonym is used to assure the identity anonymity of VTs in virtual spaces as well. As shown in Fig. \ref{scheme}, VMU and VT pseudonyms are stored in VMU pseudonym pools and VT pseudonym pools within RSUs, respectively. Our proposed scheme uses varying VMU and VT pseudonyms to conceal the real identities of VMUs and VTs. However, if the pseudonyms are single-use, i.e., VMUs and VTs discard the old pseudonyms after pseudonym changes, the pseudonyms will quickly run out. Once pseudonyms in pools are exhausted, the CA needs to allocate pseudonyms again, which leads to high pseudonym generation and communication overhead. To tackle this challenge, the proposed scheme adopts a blockchain-based VMU-VT dual pseudonym management approach, where shuffling operations enable the reuse of both VMU and VT pseudonyms \cite{johar2021proofBC}. Firstly, the source RSUs distribute VMU and VT pseudonyms to VMUs and VTs. Moreover, our proposed scheme encompasses four corresponding modules to defend against the four kinds of privacy attacks. More details are described as follows:

\begin{itemize}
    \item \textbf{\emph{Module 1: Mutual authentication}:} \textit{The mutual authentication module can defend against V2T attacks}. We consider that $VMU_1$ and $VT_1$ first receive a VMU pseudonym set \{$PID_{VMU_1}^w$\} and a VT pseudonym set \{$PID_{VT_1}^u$\}, respectively. Here $u$ is the number of VT pseudonyms. Then, $VMU_1$ and $VT_1$ initiate the mutual authentication process, where they verify the pseudonym of their counterpart in \{$PID_{VT_1}^u$\} and \{$PID_{VMU_1}^w$\}, respectively\cite{xu2021efficient}. When completing the mutual authentication, both $VMU_1$ and $VT_1$ obtain a shared secret key and create a secure communication channel for subsequent data transmission\cite{xu2021efficient}. In this case, $VT_1$ receives physical sensing data only uploaded by $VMU_1$ while $VMU_1$ receives feedback only sent from $VT_1$. Therefore, external attackers can neither capture data from $VMU_1$ nor send erroneous data to $VT_1$.
    
    \item \textbf{\emph{Module 2: VT blacklist}:} \textit{The VT blacklist module can defend against V2T attacks}. Malicious VTs often impersonate legitimate VTs to perpetrate misbehaviors such as defrauding, or sharing fake news. However, malicious VTs can be accused by legitimate VTs and then reported by RSUs to CA\cite{johar2021proofBC}. By examining the evidence in reports and historical requests of pseudonym changes in log files, the CA is competent to evaluate the validity of these reports. If the reported behaviors are genuine, the CA will revoke the use of malicious VTs' pseudonyms while revealing their true identities to all VTs in the virtual space\cite{johar2021proofBC}. Finally, the malicious VTs will be added to the VT blacklist to prevent them from interacting with legitimate VTs. 
    
    \item \textbf{\emph{Module 3: Group pseudonym change for VMUs}:} \textit{The group pseudonym change for VMUs module can defend against V2V attacks}. After driving for a while, privacy levels of VMUs decrease to the anticipated threshold. Thus, VMUs decide to change new VMU pseudonyms for improving privacy levels. We consider that attackers have eavesdropped safety messages with {$PID_{VMU_1}^1$} broadcast by the vehicle of $VMU_1$. The $VMU_1$ chooses to change the pseudonym in a social hot spot (e.g., a busy intersection), where more legitimate VMUs jointly change their VMU pseudonyms with a higher frequency for enhancing the overall privacy level \cite{yu2015mixgroup}. Then, $VMU_1$ replaces the pseudonym with {$PID_{VMU_1}^2$}. Since masses of nearby VMUs in the group that change pseudonyms together have similar features (e.g., locations and velocity), attackers will lose track of the target $VMU_1$.

    \item \textbf{\emph{Module 4: Group pseudonym change for VTs}:} \textit{The group pseudonym change for VTs module can defend against T2VSP attacks}. With the aid of VT groups, our scheme has a positive effect on defending against malicious VSPs in the virtual space. Specifically, $VT_1$ deployed in the source RSU first utilizes VT pseudonym {$PID_{VT_1}^1$} to request location-based metaverse services. To maintain seamless experiences for $VMU_1$, $VT_1$ is migrated from the source RSU to the destination one. Meanwhile, $VT_1$ is qualified to join a VT group formed on the destination RSU where legitimate VTs within the communication coverage assemble for collective pseudonym changes, and then changes its pseudonym from {$PID_{VT_1}^1$} to {$PID_{VT_1}^2$} together with other members' changing in the group. In this scenario, malicious VSPs will lose the target $VT_1$.
\end{itemize}

After pseudonym changes, both VMUs and VTs return used or expired pseudonyms to corresponding pseudonym pools in the destination RSUs when the pseudonyms stored in sets are about to run out \cite{johar2021proofBC}. Furthermore, these recycled pseudonyms are shuffled and allocated by distributed consensus (e.g., Proof-of-Pseudonym\cite{johar2021proofBC}) among different RSUs for reuse. Specifically, as RSUs are confidential and authorized, consortium blockchains\cite{cao2022blockchain} can be leveraged to ensure the security of pseudonym management and distribution relying on encryption technologies and consensus algorithms. The pseudonym shuffling transactions are packed into the blocks (i.e., distributed ledgers) among RSUs, guaranteeing the immutability and integrity of both VMU and VT pseudonyms. Therefore, the blockchain-based VMU-VT dual pseudonym management approach contributes to identity traceability and accountability in vehicular metaverses whenever a dispute or a report occurs.

\begin{figure}[t]
\centering{\includegraphics[width=0.48\textwidth]{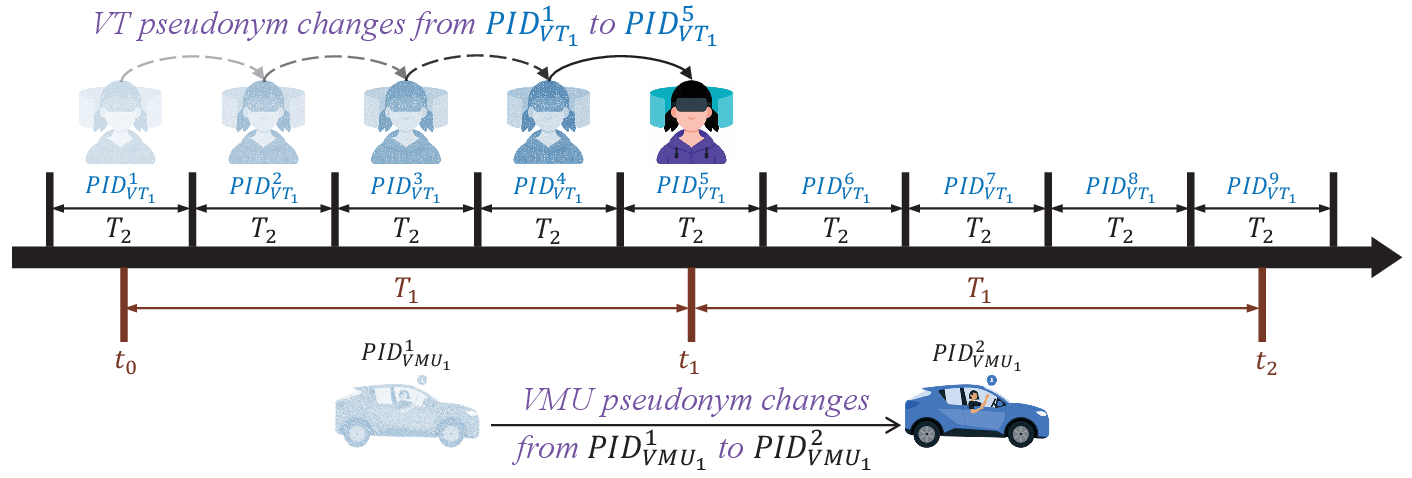}}
\caption{VMU-VT linkage mapping threat.}\label{timeline}
\end{figure}

\section{Linkage Mapping Threat and Synchronous Pseudonym Change Framework}

\subsection{VMU-VT Linkage Mapping Threat}
Although the VMU-VT dual pseudonym scheme plays a significant role in defending against the attacks, there still exists a latent safety hazard that leads to severe location privacy breaches. Specifically, we further consider an underlying threat resulting from asynchronous VMU-VT pseudonym changes. The attackers may eavesdrop safety messages from the vehicles of target VMUs in the physical space while stealing sensitive information including VT pseudonyms from target VTs in virtual spaces. By analyzing spatio-temporal background information of both, the attackers can establish mapping relationships between the identities of target VMUs and VTs \cite{kang2016location}. As VMUs and VTs change pseudonyms asynchronously, the attackers can re-identify the target by linking VMU pseudonyms with VT pseudonyms, which is called \textit{VMU-VT linkage mapping threat} in this article.

As shown in Fig. \ref{timeline}, we use a timeline of pseudonym changes to describe the threat in detail. Without loss of generality, attackers are prone to launch attacks in vehicular metaverses, because they are restricted by spatial locations in the physical world but easy to access boundless virtual spaces anywhere. Additionally, since VTs are deployed in RSUs while VT pseudonym pools are also stored in RSUs, changing VT pseudonyms incurs lower communication overhead and is more cost-effective than changing VMU pseudonyms. Therefore, we consider that $VT_1$ changes its pseudonyms four times as often as $VMU_1$ changes within a certain time period to reduce the risk of being tracked. Besides, for ease of expression, we consider that both $VMU_1$ and $VT_1$ change their pseudonyms evenly, namely with $VMU_1$ changing VMU pseudonyms every time period $T_1$ while $VT_1$ changing VT pseudonyms every time period $T_2$\cite{kang2016location}.

Here we present a concrete example to introduce the VMU-VT linkage mapping threat in Fig. \ref{timeline}. We consider that attackers have observed the pseudonym of target $VMU_1$ (i.e., $PID_{VMU_1}^{1}$) by eavesdropping safety messages before $t_0$. During the first time period $T_1$ (i.e., from $t_0$ to $t_1$), $VT_1$ changes its pseudonym sequentially from $PID_{VT_1}^{1}$ to $PID_{VT_1}^{5}$, while $PID_{VMU_1}^{1}$ remains unchanged. In a limited road network, the location-related features of VMUs and VTs (e.g., geographical locations and surrounding landscape) partially overlap. By analyzing these common features, attackers can establish a mapping relationship between VMU pseudonyms and VT pseudonyms (i.e., $PID_{VMU_1}^{1}$ corresponding to \{$PID_{VT_1}^{1}$, $PID_{VT_1}^{2}$, $PID_{VT_1}^{3}$, $PID_{VT_1}^{4}$, $PID_{VT_1}^{5}$\}), allowing them to track the target $VT_1$\cite{kang2016location}. Likewise, even though $VMU_1$ replaces its pseudonym with $PID_{VMU_1}^{2}$ at $t_1$, the VT pseudonym $PID_{VT_1}^{5}$ stays invariable. As attackers already know the correspondence between $PID_{VMU_1}^{1}$ and $PID_{VT_1}^{5}$, they can easily re-identify the target $VMU_1$ by building a mapping relationship between $PID_{VT_1}^{5}$ and \{$PID_{VMU_1}^{1}$, $PID_{VMU_1}^{2}$\}. Therefore, the attackers can keep track of the target by continually linking VMU pseudonyms with VT pseudonyms, such as linking $PID_{VMU_1}^{2}$ with \{$PID_{VT_1}^{5},\ldots, PID_{VT_1}^{9}$\}. As long as the VMU pseudonyms and VT pseudonyms are changed asynchronously, strong attackers can always follow up their targets precisely \cite{kang2016location}, resulting in serious location privacy disclosures of both VMUs and VTs in vehicular metaverses.

\begin{figure*}[t]
\centering{\includegraphics[width=0.76\textwidth]{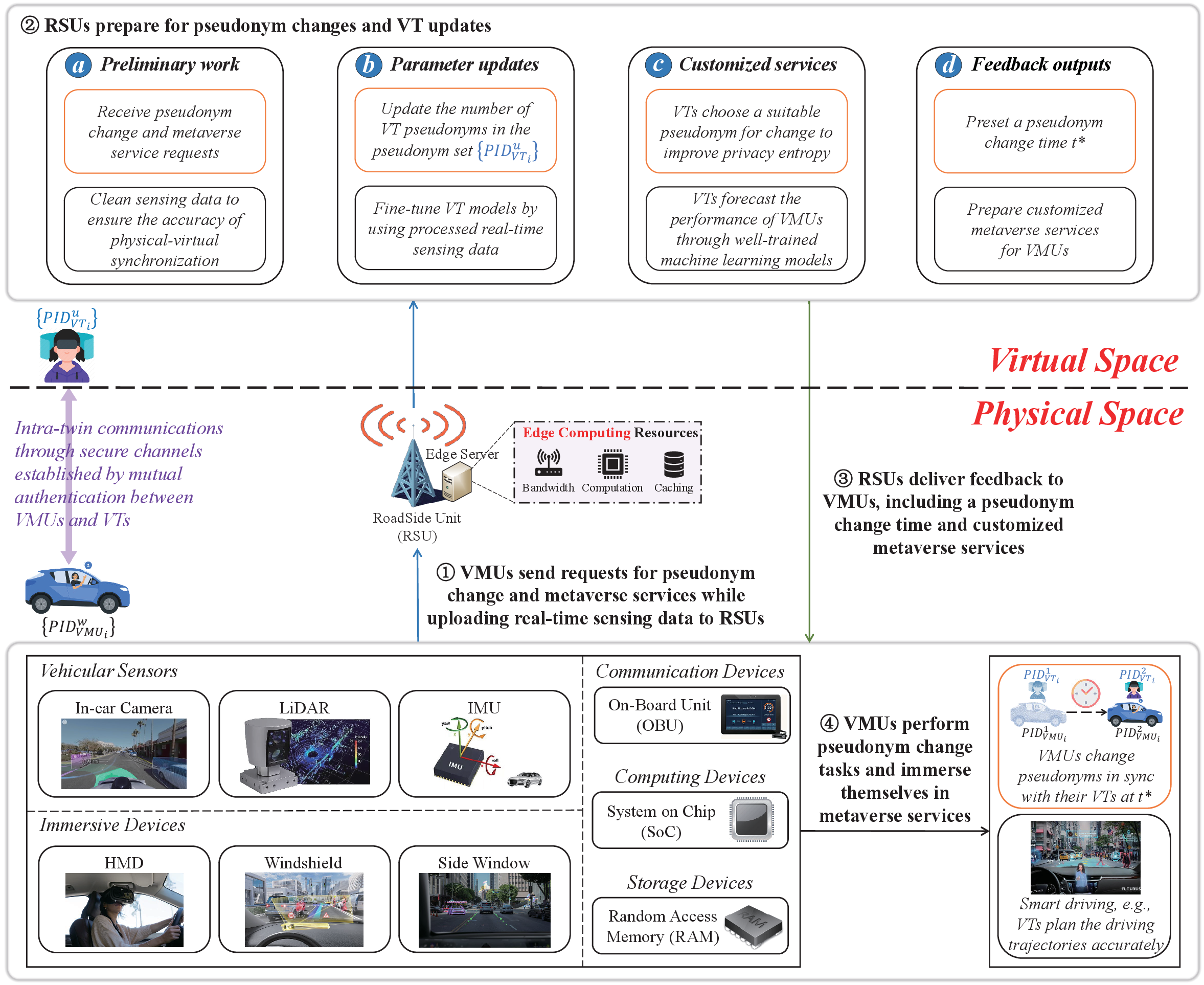}}
\caption{The synchronous VMU-VT pseudonym change framework. Both changing pseudonyms and experiencing metaverse services are realized by intra-twin communications.}\label{framework}
\end{figure*}

\subsection{Synchronous VMU-VT Pseudonym Change Framework}
To address this threat, we propose a synchronous VMU-VT pseudonym change framework based on intra-twin communications (i.e., data synchronization between VMUs and VTs) \cite{wang2023DTsurvey} in Fig. \ref{framework}. Notably, VMUs are equipped with various vehicular sensors (e.g., in-car cameras, Light Detection and Ranging (LiDAR), and Inertial Measurement Unit (IMU) suits) for real-time data acquisition and immersive devices (e.g., HMDs, windshields, and side windows) for metaverse service displays. Besides, communication, computing, and storage resources within vehicles also facilitate the processes of pseudonym changes and metaverse service experiences. To promote immersion and satisfaction in vehicular metaverses, the low-latency data flowing between VMUs and VTs is vital. Empowered by global navigation satellite system receivers in vehicles, VMUs can pre-synchronize their internal clocks with the master clocks in nearby RSUs, thus realizing accurate time synchronization with their VTs in virtual spaces\cite{seijo2020Timesy}. 

Under the premise of time synchronization, VMUs can experience immersive metaverse services and conduct synchronous VMU-VT pseudonym changes via intra-twin communications, as shown in Fig. \ref{framework}. VMUs upload metaverse service requests along with real-time sensing data to VTs for updates. Afterwards, the VTs process these data and provide feedback to instruct the performances of VMUs, by which VMUs can immerse themselves in splendid metaverse services through immersive devices \cite{zhou2022vetaverse,yu2022bi}. In addition to helping VMUs enjoy metaverse services, the intra-twin communication also supports the synchronous VMU-VT pseudonym change framework. The key steps are listed as follows:

\begin{itemize}
    \item \textbf{Step 1. Initialization and request record:} $VMU_i$, which is ready to perform a synchronous VMU-VT pseudonym change, first checks whether there are available pseudonyms in the pseudonym set \{$PID_{VMU_i}^{w}$\}. If yes, the $VMU_i$ will send a VMU pseudonym change request attached with the current timestamp to the nearest RSU through a secure channel established by mutual authentication (see $\textcircled{1}$ in Fig. \ref{framework}). Then, the RSU transfers this request and the timestamp to the CA for recording. Logging this information enables CA to trace VMUs' true identities in the events of disputes or accusations in the future, thus maintaining the accountability in vehicular metaverses\cite{kang2016location}. If no, the $VMU_i$ will apply for a new pseudonym set from the nearest RSU.
    
    \item \textbf{Step 2. Preparation for synchronous pseudonym changes:} When receiving the pseudonym change request, the RSU starts preparing pseudonym changes for both VMUs and VTs (see $\textcircled{2}$ in Fig. \ref{framework}). The RSU first updates the number of VT pseudonyms in \{$PID_{VT_i}^{u}$\}. If there are no extra pseudonyms, $VT_i$ will request a new set from the RSU where it is deployed. If adequate pseudonyms are available, $VT_i$ will choose a suitable pseudonym for replacement. To ensure synchronous changes of VMU and VT pseudonyms, the RSU presets a pseudonym change time $t^*$ for $VT_i$ in the timer\cite{kang2016location}, which is also output as feedback to instruct $VMU_i$ to change pseudonyms.
    
    \item \textbf{Step 3. Synchronous pseudonym changes:} After receiving feedback from the RSU through the intra-twin communication (see $\textcircled{3}$ in Fig. \ref{framework}), the $VMU_i$ selects a proper pseudonym from the VMU pseudonym set stored in the vehicle to perform the pseudonym change task (see $\textcircled{4}$ in Fig. \ref{framework}). Under the guidance of pseudonym change time included in feedback, both $VMU_i$ and $VT_i$ synchronously change their respective pseudonyms at the predetermined time $t^*$\cite{kang2016location}.
\end{itemize}

As VMU and VT pseudonyms of targets are changed in synchronization, attackers lose the physical and virtual identities simultaneously, thus losing track of their targets. Therefore, the proposed synchronous VMU-VT pseudonym change framework can resist the VMU-VT linkage mapping threats, protecting the location privacy of legitimate participants in vehicular metaverses effectively.

\begin{figure}[t]
\centering{\includegraphics[width=0.48\textwidth]{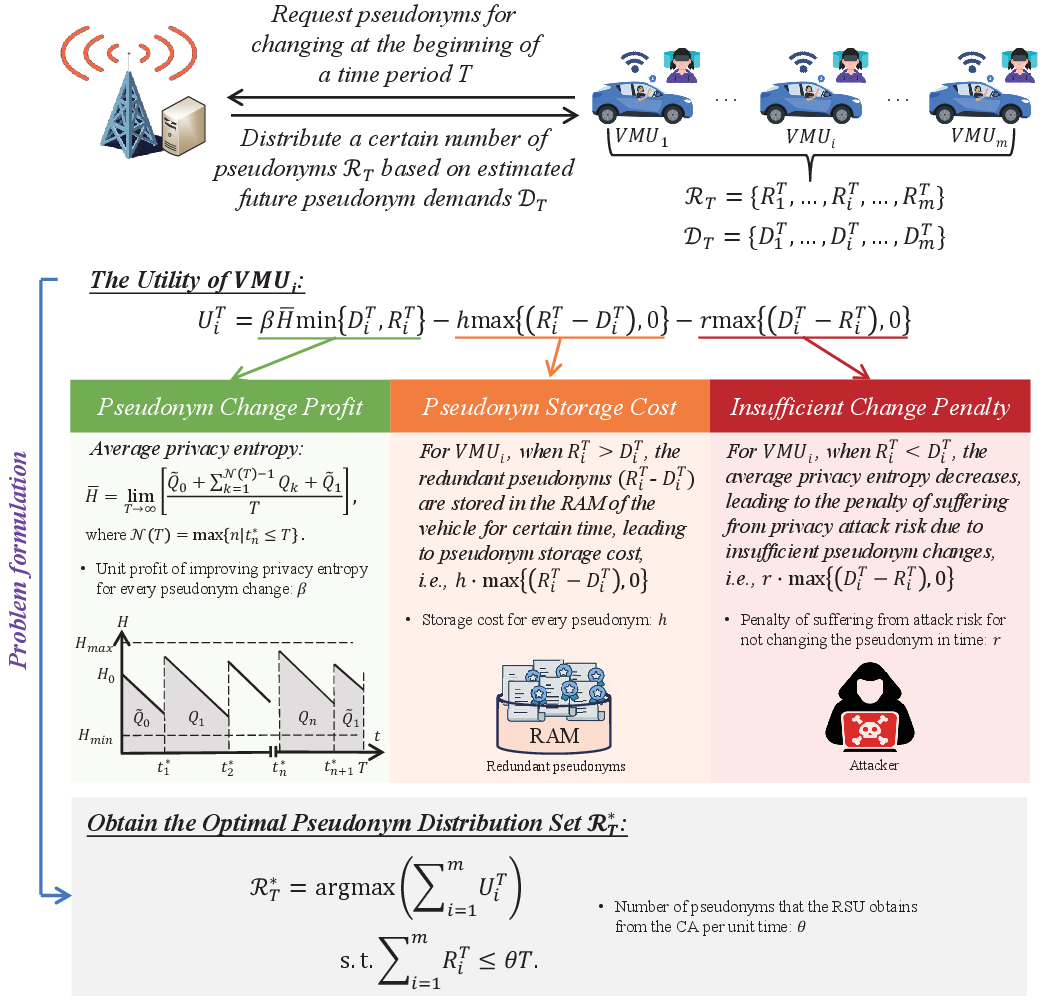}}
\caption{The pseudonym distribution problem between the RSU and VMUs based on the inventory theory.}\label{math}
\end{figure}



\begin{figure}[t]
	\centering
	\subfigure[Performance comparison between the proposed scheme and the equal distribution scheme. Note that pseudonym change frequencies of six VMUs are set to $\small\{1, 1.2, 1.4, 1.6, 1.8, 2\small\}$.]{\label{utility}\includegraphics[width=0.45\textwidth]{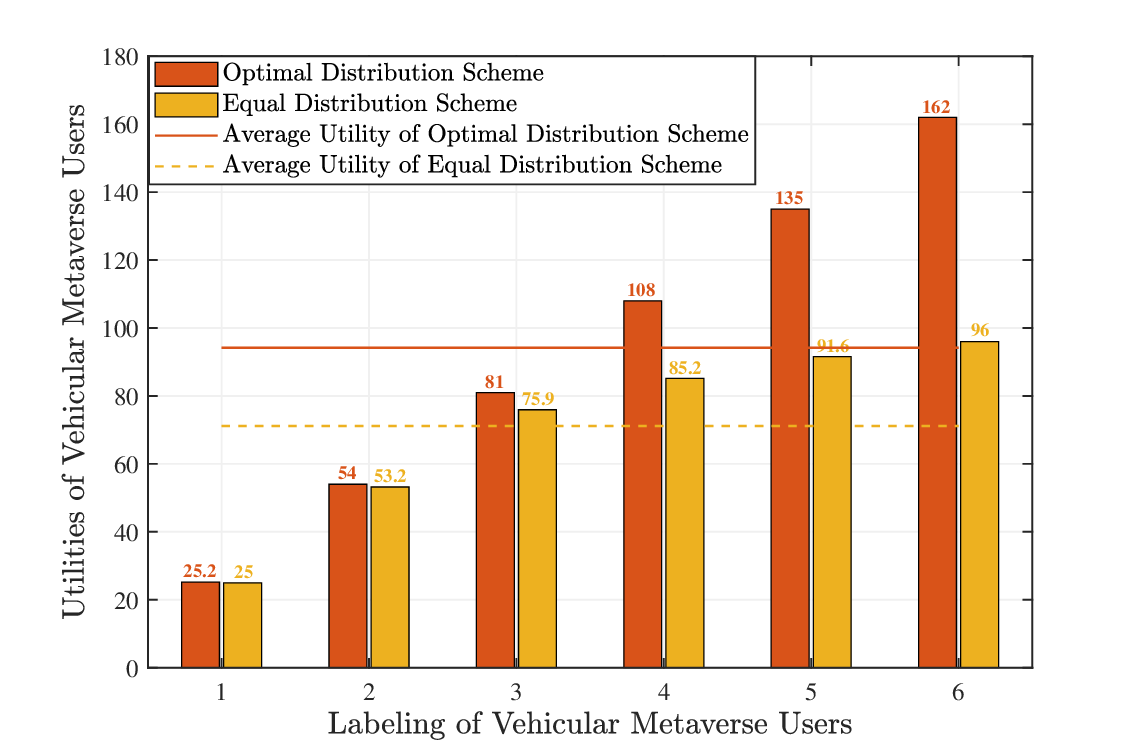}}
	\subfigure[The global utility of three VMU groups under the proposed scheme, where pseudonym change frequencies of VMU group 1, 2, and 3 are set to $\small\{1, 1.2, 1.4\small\}$, $\small\{2, 2.2, 2.4\small\}$, and $\small\{3, 3.2, 3.4\small\}$, respectively. ]
{\label{Cost_MU}\includegraphics[width=0.45\textwidth]{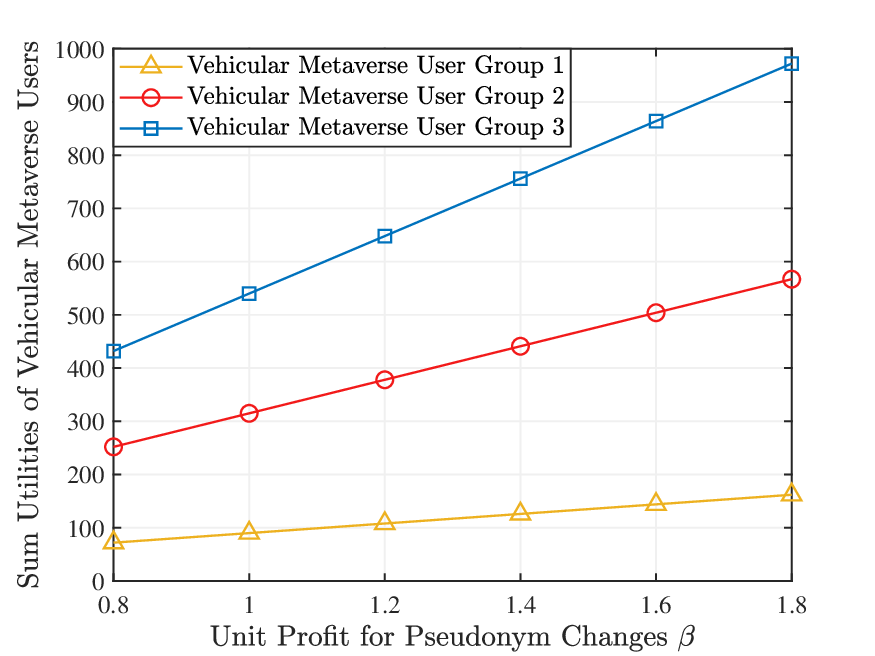}}
    \caption{The performance of the proposed on-demand pseudonym distribution scheme.}\label{performance}
\end{figure}

\section{Case Study}
In this section, we investigate a scenario where VMUs request a specific number of pseudonyms for change. We first derive average privacy entropy to quantify the increased degree of privacy protection after a pseudonym change. Then, we optimize the number of pseudonym distribution based on the inventory theory.

\subsection{Scenario Description}
As shown in Fig. \ref{math}, we consider that RSUs obtain pseudonyms from the CA at a constant rate $\theta$. At the beginning of a time period $T$, VMUs first request pseudonyms from the RSU in which their VTs are deployed. Then, the RSU distributes a certain number of pseudonyms to VMUs according to estimated future pseudonym demands of VMUs, where future pseudonym demands can be estimated by the pseudonym change frequency of VMUs based on historical observation records. After receiving pseudonyms, VMUs need to change their pseudonyms timely to ensure privacy protection.

\subsection{On-Demand Pseudonym Distribution based on Inventory Theory}
We formulate the pseudonym distribution problem between the RSU and VMUs by inventory theory. The inventory theory aims to optimize the inventory management of a business by determining the appropriate timing and quantity of orders for specific goods\cite{miriam2023inventory}. In our pseudonym distribution model, the RSU aims to develop an optimal pseudonym distribution strategy by maximizing the sum of VMU utilities.
\subsubsection{Average privacy entropy of VMUs} As an effective metric that measures the degree of privacy protection for VMUs, privacy entropy for $VMU_i$ is defined as $H_i = -\log_2 p_i$, where $p_i\in(0,1]$ is the probability of $VMU_i$ being tracked after a pseudonym change\cite{kang2016location}. We consider that the privacy entropy of $VMU_i$ decreases linearly over time with slope $\alpha$ before it reaches the minimum privacy entropy $H_{min}$. After $VMU_i$ synchronously changing pseudonyms with $VT_i$, the privacy entropy of $VMU_i$ increases to $(H_{max}-p_iH_0)$, where $H_{max}$ is the maximum privacy entropy. Therefore, the privacy entropy function over time is sawtooth. To better assess the increased degree of privacy protection after a pseudonym change, we derive the average privacy entropy $\overline{H}$ for VMUs. As shown in Fig. \ref{math}, the average privacy entropy is the area under the sawtooth function normalized by the time interval.

\subsubsection{VMU utility}
We denote $R_i^T$ as the number of pseudonyms requested by $VMU_i$ at the beginning of the time period $T$ and $D_i^T$ as future pseudonym demands in the time period $T$, respectively\cite{kang2016demand}. As shown in Fig. \ref{math}, the utility of $VMU_i$ is denoted as $U_i^T$, which consists of pseudonym change profits, pseudonym storage costs, and insufficient change penalties. Specifically,  $VMU_i$ can obtain profits from the increased degree of privacy protection after each pseudonym change. However, if $R_i^T>D_i^T$, the redundant pseudonyms have to be stored in vehicles for a certain time, leading to storage costs\cite{kang2016demand}. Note that the storage cost per pseudonym is less than the change cost per pseudonym. If $VMU_i$ cannot satisfy pseudonym demands, i.e., $R_i^T<D_i^T$, $VMU_i$ will bear the penalties of being exposed to privacy risk due to the reduction in average privacy entropy.

\subsubsection{Problem formulation}
To obtain the optimal pseudonym distribution set $\mathcal{R}_T^* = \small\{R_i^T\small\}$, we maximize the global utility $\sum_{i=1}^mU_i^T$, where $\mathcal{R}_T^*$ exists only if $\sum_{i=1}^mR_i^T\leq\theta T$. Note that the global utility function is concave, indicating that there exists the maximum value of this function, which can be solved approximately using the genetic algorithm \cite{kang2016demand}.

\subsection{Numerical Results}
To prove the efficiency of the on-demand pseudonym distribution scheme, we compare the proposed scheme with an equal distribution scheme, where pseudonyms are equally distributed to VMUs. The pseudonym change strategy complies with the ETSI TR 103 415 standard\footnote{\url{https://www.etsi.org/deliver/etsi_tr/103400_103499/103415/01.01.01_60/tr_103415v010101p.pdf}}. Specifically, we set unit time as per minute and consider that the RSU obtains $10$ pseudonyms per minute, i.e., $\theta = 10$, and suppose there exist six VMUs requesting pseudonyms from the RSU, where the process of pseudonym requests from VMUs follows the Poisson process in an observation time period\cite{kang2016demand}, set to $1$ hour. Besides, we consider that $p_i$ follows a uniform distribution in $(0.0.5]$ and factors $H_{max}$, $H_0$, $H_{min}$, $\alpha$, $h$, and $r$ are set to $1.5$, $1$, $0.25$, $1$, $0.1$, and $0.3$, respectively.

Figure \ref{utility} presents the respective utility of six VMUs under the on-demand pseudonym distribution scheme and the equal distribution scheme. Without loss of generality, the larger the serial number of VMU, the more frequently the pseudonym is changed. We can observe that, for each VMU, the utility under the proposed scheme is higher than that under the equal distribution scheme, and the average utility of VMUs under the proposed scheme is $33.8\%$ higher than that under the equal distribution scheme. The reason is that the utilization of pseudonyms is maximized by distributing pseudonyms to VMUs based on their actual demands. Figure \ref{Cost_MU} illustrates the global utility of three VMU groups under the proposed scheme, where each group consists of three VMUs. We can find that, for the fixed unit profit for pseudonym changes $\beta$, the global utility of VMU group 3 with the highest average pseudonym change frequencies is maximum, indicating that VMUs can better enhance the degree of privacy protection by changing pseudonyms in VMU groups with a higher average pseudonym change frequency.

\section{Conclusion}
In this article, we studied privacy attacks and defenses for Vehicular Twin (VT) migrations in vehicular metaverses. We systematically introduced the VT migration process and presented four kinds of specific privacy attacks compromising the identity and location privacy of both Vehicular Metaverse Users (VMUs) and VTs. To defend against these attacks, we proposed a VMU-VT dual pseudonym scheme consisting of four corresponding modules. Furthermore, we proposed a synchronous VMU-VT pseudonym change framework to address an underlying threat resulting from asynchronous pseudonym changes between VMUs and VTs. Finally, we carried out a case study to demonstrate the significant efficiency of the on-demand pseudonym distribution strategy compared with the equal distribution strategy. In the future, we will further explore the average privacy entropy model to better quantify the degree of privacy protection for pseudonym changes, and delve into the use of artificial intelligence tools (e.g., deep reinforcement learning) to optimize the pseudonym distribution in vehicular metaverses.

\bibliographystyle{IEEEtran}
\bibliography{reference}

\end{document}